\documentclass[preprint, 3p]{elsarticle}

\usepackage{lineno,hyperref}

\journal{Electrochimica Acta}

\begin{document}

\begin{frontmatter}

\title{Study on the free corrosion potential at an interface between an Al electrode and an acidic aqueous NaCl solution through density functional theory combined with the reference interaction site model}

\author{Koichi Kano$^a$}
\ead{kano.koichi@kki.kobelko.com}

\author{Satoshi Hagiwara$^b$}
\author{Takahiro Igarashi$^c$}
\author{Minoru Otani$^b$}
\ead{minoru.otani@aist.go.jp}

\address{$^a$ Kobelco research institute. INC., Takazukadai 1-5-5, Nishi, Kobe, Hyogo 651-2271, Japan}
\address{$^b$ National Institute of Advanced Industrial Science and Technology (AIST), 1-1-1, Umezono, Tsukuba, Ibaraki 305-8568, Japan}
\address{$^c$ Research Group for Corrosion Resistant Materials, Japan Atomic Energy Agency (JAEA), Tokai, Japan}

\begin{abstract}
We investigated the free corrosion potential at an interface between an Al electrode and an aqueous NaCl solution (NaCl(aq)) under acidic conditions via density functional theory combined with the effective screening medium and reference interaction site model (ESM-RISM). 
Firstly, the electrode potentials for the anodic and cathodic corrosion reactions were obtained from the grand potential profile as a function of the electron chemical potential at the interface. 
Thereafter, we determined the free corrosion potential using the Tafel extrapolation method. 
The results of the free corrosion potential were consistent with previous experimental data. 
By controlling the pH of the NaCl(aq), we determined the pH dependence of the free corrosion potential, and the results reasonably reproduce the previous experimental results. 
Our results indicated that the ESM-RISM method duly described the environmental effect of an acidic solution and precisely determined the free corrosion potential. 
Therefore, the application of this method presents an efficient approach toward calculating the free corrosion potential for various reactions.
\end{abstract}


\end{frontmatter}


\section{Introduction}\label{s:intro}

Protecting materials from corrosion, which entails charge-transfer (CT) reactions at the material/electrolyte interface, plays an essential role in controlling their lifespan \cite{bethencourt1998lanthanide, kritzer2004corrosion, chen2011review, verma2017corrosion}.  
During corrosion reactions, materials react with their environment, and thereafter dissolve into an electrolyte solution. 
Therefore, owing to the corrosion reaction, the aging degradation of the material gradually progresses. 
For corrosion protection, the free corrosion potential, which is an equilibrium potential of the corrosion reaction, is a significant physical quantity because it is an indicator of the corrosion reaction from an electrochemical perspective. 
Consequently, the free corrosion potential of various materials has been extensively measured \cite{dibari1971electrochemical, minakawa19841, simoes2008assessment, mansouri2011anodic, sherif2011effects, acosta2014power, mondal2016development}. 

Theoretically, the corrosion reaction has been investigated through first-principles density functional theory (DFT) studies \cite{ropo2007theoretical, taylor2008first, ropo2011first, ma2017first, ke2019density}. 
However, there are merely a few studies on the free corrosion potential from DFT calculations \cite{ma2017first}, as it strongly depends on the environmental effects at the material/electrolyte interface, such as temperature, pH , and halogen species of the electrolyte. 
Therefore, the development of a simulation technique that accurately describes the electrochemical environment at the material/electrolyte interface is essential. 

First-principles electronic structure calculations combined with first-principles molecular dynamics (FPMD) is a powerful tool for studying the physical phenomena at the electrochemical interface \cite{otani2008structure, ikeshoji2012charged, ikeshoji2017toward, ambrosio2018ph, wiktor2019electron}. 
However, the FPMD approach is expensive as it due to high computational costs because numerous sampling trajectories are needed to describe the average thermal properties of the solution. 
In the free corrosion potential, the local cell generates a potential difference between the anodic and cathodic reactions. 
The difference introduces excess electrons on the surface, and the ions in the electrolyte solution screen the electrostatic field generated by the excess electrons, resulting in the formation of an electric double layer (EDL). 
Therefore, the whole system is neutrally charged, but the origin of the compensating charge on the electrode and in the electrolyte differs. 
The former is the electron in an electrode, while the latter represents the ions in an electrolyte solution. 
To introduce these additional charges into the system, we need to use a simulation technique that enables the grand canonical treatment of both electrons and ions. In FPMD simulation, such a particle treatment approach is challenging. Thus, a more flexible simulation technique to describe electrochemical interfaces is required.

To overcome such FPMD limitations, a hybrid solvation model, which is the DFT method combined with the classical liquid or classical DFT, has been developed \cite{mathew2014implicit, sakong2015density, lespes2015using, sundararaman2017evaluating}. In the hybrid solvation model, we explicitly treat an electrode and a few molecules in the electrolyte. Conversely, the remaining electrolyte is replaced by the classical continuum model. The hybrid solvation model has been applied to the electrode/electrolyte interface and satisfactorily describes differential capacitances and electrode potentials \cite{mathew2014implicit, lespes2015using, sundararaman2017evaluating}.

In this study, the DFT method combined with the effective screening medium (ESM) technique \cite{otani2006first, hamada2013improved} and the reference interaction site model (RISM) \cite{hirata1981extended, hirata1982application, kovalenko1998three}, termed as ESM-RISM \cite{nishihara2017hybrid, haruyama2018electrode}, is applied to the Al/aqueous NaCl (NaCl(aq)) solution interface. 
As Al is widely used in our daily lives because of its high anticorrosion properties, it is one of the essential materials applied in the field of corrosion. 
Here, we focus on the corrosion reaction under acidic conditions.
The objective of this study is to calculate the free corrosion potentials at the interface between the Al electrode and NaCl(aq) under acidic conditions using the ESM-RISM approach.

This paper is organized as follows: 
Firstly, the ESM-RISM method is briefly elucidated in section 2, and we describe a series of corrosion reactions used in this study. 
Next, we explain the method for calculating the equilibrium potential from electrochemical and thermodynamic perspectives.
Thereafter, the Tafel equation, using ESM-RISM to calculate the free corrosion potential, is discussed. 
Section 3 presents the results of the calculations. 
Here, we first present a series of results of the free corrosion potentials at the Al/NaCl(aq) interfaces electrochemically calculated using the ESM-RISM method. Next, the differences in the results between the electrochemical and thermodynamic methods are examined. 
We thereafter compare the pH dependence of the free corrosion potentials between the present theory and the previous experiment. 
Finally, the results of the free corrosion potentials depending on the surface orientation are revealed.
In section 4, we discuss the results of the present calculations from the perspectives of methodology, pH, and surface dependence of the corrosion potentials. 
The last section provides a summary of the present study.

\section{Methods and models} \label{s:method}

In this section, we describe the theoretical and computational details used in this study. 
Firstly, we briefly describe the formulation of the ESM-RISM method and grand potential. 
Secondly, a set of reactions that is assumed to occur under corrosive conditions is described. 
Next, we explain the calculation methods for the electrode potential, current, and free corrosion potential for the anodic and cathodic reactions. 
Finally, the computational details are provided. 

\subsection{ESM-RISM method}
Firstly, we briefly describe the ESM-RISM method \cite{nishihara2017hybrid, haruyama2018electrode}.
The ESM-RISM is a hybrid solvation model, which is a self-consistent electronic structure calculation that considers the interactions between explicit particles and implicit solvent molecules (and ions) in a solution. 
The ESM-RISM can treat not only electrons in the explicit system \cite{bonnet2012first} but also RISM components in the implicit solution under grand canonical conditions \cite{nishihara2017hybrid}. 
Consequently, the well-defined inner potential $\Phi_{\mathrm{S}}$, which is the bulk potential of the solution, is used as the reference electrode potential \cite{haruyama2018electrode}.

In the ESM-RISM approach, the total energy of the whole system is described by the Helmholtz free energy $A$ as
\begin{equation}
A=E_{\rm DFT} + \Delta \mu_{\rm solv}, 
\end{equation}
where, $E_{\rm DFT}$ and $\Delta \mu_{\rm solv}$ are the total energy of the DFT and the excess chemical potential of the solution determined by the RISM, respectively. 
$\Delta \mu_{\rm solv}$ corresponds to the excess free energy by considering the distribution of the implicit solvent at a given temperature, which is determined by solving RISM equations using a closure function \cite{hirata1981extended, hirata1982application, kovalenko1998three}.

We define the grand potential $\Omega$ using $A$ by:
\begin{equation}
\Omega = A - \mu_{\rm e}(N_{\rm e}-N^0_{\rm e}) = A - \mu_{\rm e}\Delta N_{\rm e}, \label{grand}
\end{equation}
where $N_{\rm e}$ and $N^0_{\rm e}$, respectively, denote the total number of electrons for the system with and without an excess charge. 
In the formulation of $\Omega$, we consider only $\mu_{\rm e}$ and $\Delta N_{\rm e}$ because $\Delta \mu_{\rm solv}$ includes a change in the chemical potential and the number of RISM particles by solving the RISM equations. 
We expect that the $\Omega$ profiles are the approximate inverse of the parabolic shape centered at the potential of zero charge (PZC), which is consistent with the constant capacitance model, as shown in previous studies \cite{haruyama2018electrode, weitzner2020toward}. 

According to the previous study\cite{weitzner2020toward}, a polynomial model is suitable for describing the profile of $\Omega$. We expand the $\Omega$ around the potential at $\Phi_{0}$ to the third order, as follows: 
\begin{eqnarray}
\Omega(\Phi) = \Omega_{0} &+& \frac{1}{1!} \left. \frac{d\Omega}{d\Phi} \right |_{\Phi_0} ( \Phi - \Phi_0 ) \nonumber \\
&+& \frac{1}{2!} \left. \frac{d^2\Omega}{d\Phi^2} \right |_{\Phi_0} (\Phi - \Phi_0 )^2 \nonumber \\
&+& \left. \frac{1}{3!} \frac{d^3\Omega}{d^3\Phi} \right |_{\Phi_0} (\Phi - \Phi_0 )^3 \nonumber \\
&+& \mathcal{O} \{ (\Phi - \Phi_0 )^4 \}. 
\end{eqnarray}
Here, the notation of the first term denotes the $\Omega(\Phi_0)$, while the second, third, and fourth coefficients correspond to the contributions to the $\Omega$ from the excess surface charge, capacitance, and change in the capacitance by the change in the potential, respectively. 
The electrode potential $\Phi$ is obtained by using the electron chemical potential $\mu_{\rm e}$ as $\Phi=-\mu_\mathrm{e}/e$. We will use this polynomial-fitted $\Omega$ in the analysis of the potentials.

\subsection{Corrosion reaction}

\begin{figure} [htb]
\centering\includegraphics[width=65mm]{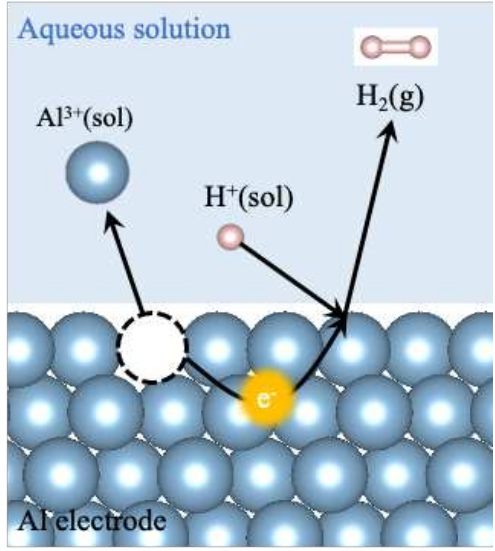}
\caption{\label{corr}
(Color online). Schematic of the corrosion reaction under acidic conditions of the aqueous solution. In the anodic reaction, Al$^{3+}$ ions are dissolved into the solution, leaving electrons at the electrode surface, while in the cathodic reaction, the proton captures the surface electrons and becomes a gas phase H$_2$ molecule. The cyan and pink balls denote Al and H atoms, respectively. The blue-shaded area represents the region of an aqueous solution.}
\end{figure}

Figure~\ref{corr} shows the schematic of the corrosion reaction in acidic aqueous solutions. 
The corrosion reaction is considered as two CT reactions, \textit{i.e.}, anodic and cathodic reactions. 
For Al, the anodic reaction for the acidic aqueous solution is written as follows:
\begin{eqnarray}
\! \! \! \! \!  \mathrm{Al(s)} &\rightarrow& \mathrm{Al^{3+}(sol)} + 3e^-(\mathrm{M}). \label{a0}
\end{eqnarray}
Here, we consider that the anodic reaction is Al in the solid state (Al(s)) dissolving into the electrolyte as Al$^{3+}$ ion (Al$^{3+}$(sol)) remaining electrons on the Al surface ($e^-$(M)). 
When the electrolyte solution is acidic NaCl(aq), the Cl$^-$ ions participate in the anodic reaction as follows \cite{verma2017corrosion}:
\begin{eqnarray}
 \! \! \! \! \!  \mathrm{Al(s)} &+& \mathrm{3Cl^-(sol)} \nonumber \\
 & \rightarrow & \mathrm{AlCl_3(sol)                    } + 3e^-(\mathrm{M}), \\
 \! \! \! \! \!  \mathrm{Al(s)} &+& \mathrm{3Cl^-(sol)} \nonumber \\
 & \rightarrow & \mathrm{\frac{1}{2}Al_2Cl_6(sol)} + 3e^-(\mathrm{M}), \\
 \! \! \! \! \!  \mathrm{Al(s)} &+& \mathrm{4Cl^-(sol)}  \nonumber \\
 & \rightarrow & \mathrm{AlCl_4^-(sol)                 } + 3e^-(\mathrm{M}),
\end{eqnarray}
where we consider the complex formation resulting from the Al$^{3+}$ combined with the Cl$^{-}$ ion of NaCl(aq). 
Since we focus on the corrosive reaction under the acidic NaCl(aq), the hydrate complex of Al$^{3+}$ (\textit{e.g.} Al(H$_2$O)$_x^{3+}$, Al(OH)(H$_2$O)$_x^{2+}$, Al(OH)$_2$(H$_2$O)$_x^{+}$, and so on) is not explicitly considered in the present study. Thus, we consider the hydration effect on the Al$^{3+}$ within the RISM theory. 
In the neutral and basic pH, the hydrate complex of Al$^{3+}$ mainly participates to the corrosion reaction and formation of the passivation layer \cite{yasuda1990pitting}. 
The cathodic reaction entails the protons in the aqueous solution receiving electrons at the electrode surface, and then, the H$_2$ molecule in the gas phase (H$_2$(g)) is produced, as shown in figure~\ref{corr}. 
Thus, the cathode reaction is defined as: 
\begin{eqnarray}
\! \! \! \! \! 2\mathrm{H}^+(\mathrm{sol}) + 2e^-(\mathrm{M}) \rightarrow \mathrm{H_2(g)}. \label{c0}
\end{eqnarray}

Here, we modeled that the corrosion reaction occurs at the electrode/electrolyte interface, and Al$^{3+}$ dissolves from the surface into the NaCl(aq) with leaving the surface vacancy. 
Consequently, as the practical forms of a set of corrosion reactions (\ref{a0})--(\ref{c0}), we consider that the dissolved Al$^{3+}$ ion (or complex) originates from the electrode surface, and the vacancy-type defect (V$_{\mathrm{Al}}$) is formed on the Al surface by the anodic reaction. In addition, we normalize the number of transferred electrons to one for the corrosion reactions. Therefore, we rewrite the anodic and cathodic reactions as follows: 
\begin{eqnarray}
\! \! \! \! \! 1/3 \mathrm{Al}_{m}(\mathrm{s}) & \rightarrow & 1/3 \mathrm{Al^{3+}(sol)} \nonumber \\
&+& 1/3 \mathrm{Al}_{m-1}(\mathrm{s}) + e^-(\mathrm{M}), \label{a3} \\
\! \! \! \! \! 1/3 \mathrm{Al}_{m}(\mathrm{s}) &+& \mathrm{Cl^-(sol)} \rightarrow 1/3 \mathrm{AlCl_3(sol)  } \nonumber \\
&+& 1/3 \mathrm{Al}_{m-1}(\mathrm{s}) + e^-(\mathrm{M}), \label{ac3} \\
\! \! \! \! \! 1/3 \mathrm{Al}_{m}(\mathrm{s}) &+& \mathrm{Cl^-(sol)} \rightarrow 1/6 \mathrm{Al2Cl_6(sol) } \nonumber \\
&+& 1/3 \mathrm{Al}_{m-1}(\mathrm{s}) + e^-(\mathrm{M}), \label{a2c6} \\
\! \! \! \! \! 1/3 \mathrm{Al}_{m}(\mathrm{s}) &+&     4/3 \mathrm{Cl^-(sol)} \rightarrow 1/3 \mathrm{AlCl_4^-(sol)} \nonumber \\
&+& 1/3 \mathrm{Al}_{m-1}(\mathrm{s}) + e^-(\mathrm{M}), \label{ac4} \\
\! \! \! \! \! \mathrm{H_3O}^+(\mathrm{sol}) &+& e^-(\mathrm{M}) \nonumber \\
&\rightarrow& 1/2 \mathrm{H_2(g)} + \mathrm{H_2O (sol)}. \label{cc}
\end{eqnarray}
Here, $m$ denotes the number of surface Al atoms. 
To determine the equilibrium and free corrosion potentials, we use equations (\ref{a3})--(\ref{ac4}) and (\ref{cc}) for the anodic and cathodic reactions, respectively. In the previous study, Haruyama et al.~calculated H$_2$ generation potential at the Pt(111)/1M HCl(aq) interface using reaction~(\ref{cc}), which agrees well with the experimental standard hydrogen electrode (SHE) potential by considering contact (or Volta) potential difference \cite{haruyama2018electrode}. 

\subsection{Equilibrium potential}
To evaluate the free corrosion potential, we determined the equilibrium potentials for the anodic and cathodic reactions using the ESM-RISM method. 
There are two pathways for calculating the equilibrium potential ($E_{\rm eq}$): (i) $E_{\rm eq}$ is determined by $\mu_{\mathrm{e}}$ at the crossing point of $\Omega$ for the left- and right-hand components of the reactions, which gives the equilibrium chemical potential of the electron ($\mu_{\mathrm{eq}}$); and (ii) $E_{\rm eq}$ is calculated by the formulation of electromotive force (EMF) with the difference in Gibbs free energies between before and after the reactions ($\Delta G$). Methods (i) and (ii) are, respectively, written as follows:
\begin{eqnarray}
\! \! \! \! \! \! \! \! \! \! \! \! (\mathrm{i}) \ \ E^{\mathrm{a(c)}}_{\rm eq} \ [\mathrm{vs.} \ \Phi_{\mathrm{S}}] &=& -\mu^{\mathrm{a(c)}}_{\rm eq} [\mathrm{vs.} \ \Phi_{\mathrm{S}}]/e, \label{ec} \\
\! \! \! \! \! \! \! \! \! \! \! \! (\mathrm{ii}) \ \ E^{\mathrm{a(c)}}_{\rm eq} \ [\mathrm{vs.} \ \Phi_{\mathrm{S}}] &=& -\Delta G^{\mathrm{a(c)}} /n^{\mathrm{a(c)}}F, \label{thermo} \\
\Delta G^{\mathrm{a(c)}} &=& G^{\mathrm{a(c)}}_{\mathrm{L(R)}} - G^{\mathrm{a(c)}}_{\mathrm{R(L)}}. \label{dG}
\end{eqnarray}
Here, superscripts ``a'' and ``c'' indicate physical quantities for anodic and cathodic reactions, respectively, throughout this paper. 
The subscript ``L(R)'' represents the left- (right-)hand component of the reaction. 
Furthermore, $n$, and $F$ indicate the number of reacted electrons and the Faraday constant, respectively. 
As using Eq.~(\ref{ec}) corresponds to the $E_{\rm eq} $ determined by controlling the bias potential, we refer to method-(i) as the \textit{electrochemical method}. In contrast to the electrochemical method, we denote method-(ii), using Eqs.~(\ref{thermo}) and~(\ref{dG}), as the \textit{thermodynamic method}. 
In the following two subsections, we describe details of the electrochemical and thermodynamic methods used in the present study. 

\subsubsection{Electrochemical method}
Firstly, we describe the practical form of the electrochemical method using reactions~(\ref{a3}) and (\ref{cc}) as those for the anode and cathode, respectively. 
In this method, we calculate the $\mu_{\mathrm{eq}}$ from $\mu_{\mathrm{e}}$ at the crossing point of the $\Omega$ profiles for the left- and right-hand components of the reactions. The grand potentials for the anodic and cathodic reactions ($\Omega^{\mathrm{a}}$ and $\Omega^{\mathrm{c}}$) are defined as follows: 
\begin{eqnarray}
\Omega^{\mathrm{a}}_\mathrm{L} &=& 1/3\Omega( \mathrm{Al}_m(\mathrm{s})   ), \\
\Omega^{\mathrm{a}}_\mathrm{R} &=& 1/3\Omega( \mathrm{Al}_{m-1}(\mathrm{s}) + \mathrm{Al}^{3+}(\mathrm{sol}) ), \\
\Omega^{\mathrm{c}}_\mathrm{L} &=& \Omega( \mathrm{Al}_m(\mathrm{s}) + \mathrm{H_3O^+(sol)} ) \nonumber \\
&+& E_{\mathrm{ZP}}(\mathrm{H_3O^+(sol)}), \\
\Omega^{\mathrm{c}}_\mathrm{R} &=& \Omega( \mathrm{Al}_m(\mathrm{s}) + \mathrm{H_2O(sol)}  ) \nonumber \\
&+& E_{\mathrm{ZP}}(\mathrm{H_2O(sol)}) \nonumber \\
&+& 1/2 \{ A(\mathrm{H_2(g)})+E_{\mathrm{ZP}}(\mathrm{H_2(g)}) \}.
\end{eqnarray}
Subscripts ``L'' and ``R'' denote the left- and right-hand components of the anodic and cathodic reactions, respectively. 
$E_{\rm ZP}$ is the zero-point vibration energy of isolated molecules or ions. 
Moreover, $A[\mathrm{H_2(g)}]$ represents the sum of the DFT energies for the $\mathrm{H_2}$ molecule in an isolated system and the standard molar entropy term $-TS$ ($=-298.15\mathrm{K} \times 130.6\mathrm{Jmol^{-1}K^{-1}}=0.404\mathrm{eV}$ \cite{databese}) for the $\mathrm{H_2}$ gas. 
Contrasting to the gas phase calculation, $-TS$ term for the solvated molecular or ionic systems was included in the $\Delta \mu_{\rm solv}$ within the ESM-RISM theory \cite{johnson2016small}. 
As the cathodic reaction indicates the CT reaction via the surface, we include the surface term for the formulation of $\Omega$. 
The electrochemical method correctly describes the change in the EDL at the interface caused by the electrode potential. 
Therefore, we naturally describe the change in the interfacial capacitance caused by defect formation at the electrode surface. 
Nevertheless, because the interaction between the electrode surface (M) and solvated ions (or molecules) (I) is weak, $\Omega$ is separated into the M and I parts as follows:
\begin{eqnarray}
\Omega(\mathrm{M+I}) &=& \Omega(\mathrm{M}) + \Omega(\mathrm{I}). 
\end{eqnarray}
Notably, these separation forms are not applicable to desolvated ion problems, such as the charge/discharge process of the lithium-ion battery \cite{haruyama2018analysis}. 
We rewrite the formulations for $\Omega^{\mathrm{a}}$ and $\Omega^{\mathrm{c}}$ as follows:
\begin{eqnarray}
\Omega^{\mathrm{a}}_\mathrm{L} &=& 1/3\Omega( \mathrm{Al}_m(\mathrm{s})     ), \label{oal} \\
\Omega^{\mathrm{a}}_\mathrm{R} &=& 1/3\Omega( \mathrm{Al}_{m-1}(\mathrm{s}) )  \nonumber \\
&+& 1/3 A( \mathrm{Al}^{3+}(\mathrm{sol}) ) -\mu_{\mathrm{e}}, \label{oar} \\
\Omega^{\mathrm{c}}_\mathrm{L} &=& \Omega( \mathrm{Al}_m(\mathrm{s}) ) + A (\mathrm{H_3O^+(sol)} ) \nonumber \\
&-& \mu_{\mathrm{e}} + E_{\mathrm{ZP}}(\mathrm{H_3O^+(sol)}), \label{ocl} \\
\Omega^{\mathrm{c}}_\mathrm{R} &=& \Omega( \mathrm{Al}_m(\mathrm{s}) ) + A (\mathrm{H_2O(sol)}   ) \nonumber  \\
&+& E_{\mathrm{ZP}}(\mathrm{H_2O(sol)})  \nonumber  \\
&+& 1/2 \{ A(\mathrm{H_2(g)})+E_{\mathrm{ZP}}(\mathrm{H_2(g)}) \}. \label{ocr}
\end{eqnarray}
Here, we use the definition of $\Omega$ described by Eq.~(\ref{grand}) for deriving the separated formula.
Practically, we use the final formula to determine the equilibrium potentials for the anodic and cathodic reactions. 

\subsubsection{Thermodynamic method}
Next, we briefly describe the thermodynamic method. We also use reactions (\ref{a3}) and (\ref{cc}) at the anode and cathode, respectively. 
Using the ESM-RISM, we calculate the electrode potential $E^{\rm a(c)}$ measured from $\Phi_{\rm S}$ with the EMF formulation. 
The ordinal formulation of EMF and Eq.~(\ref{thermo}) differ from the electrochemical reaction between the neutral and charged states. 
Thus, the former corresponds to the full-cell reaction, while the latter indicates a half-cell reaction. 
Previous DFT studies \cite{haruyama2018analysis, haruyama2019two, wang2006oxidation, grindy2013approaching} show that the EMF derived through thermodynamics is in good agreement with that of the experimental results. 
The equilibrium potential obtained by the electrochemical method agrees with that derived by the thermodynamic method, which has been shown by Haruyama et.~al \cite{haruyama2018electrode}. 
The definitions of the Gibbs free energy for the left- and right-hand components of the anodic and cathodic reactions are:
\begin{eqnarray}
G^{\mathrm{a}}_\mathrm{L} &=& 1/3 A( \mathrm{Al}_m(\mathrm{s})   ), \label{gal} \\
G^{\mathrm{a}}_\mathrm{R} &=& 1/3 A( \mathrm{Al}_{m-1}(\mathrm{s}) ) \nonumber \\ 
&+& 1/3 A( \mathrm{Al}^{3+}(\mathrm{sol}) ), \label{gar} \\
G^{\mathrm{c}}_\mathrm{L} &=& A(\mathrm{H_3O^+(sol)} ) \nonumber \\
&+& E_{\mathrm{ZP}}(\mathrm{H_3O^+(sol)}), \label{gcl} \\
G^{\mathrm{c}}_\mathrm{R} &=& A(\mathrm{H_2O(sol)}  ) + E_{\mathrm{ZP}}(\mathrm{H_2O(sol)}) \nonumber \\
&+& 1/2 \{ A(\mathrm{H_2(g)})+E_{\mathrm{ZP}}(\mathrm{H_2(g)}) \}. \label{gcr}
\end{eqnarray}
We use the Helmholtz free energy ($A$) as the approximate Gibbs free energy ($G$) because $A$ is the same as $G$ by neglecting the volume$\times$pressure term. 
When we can neglect the change in the details of the EDL, the thermodynamic method gives the same result of $E_{\mathrm{eq}}$ as that of the electrochemical. 
In such a case, the thermodynamic method is flexible compared to the electrochemical approach because the grand potential calculations require many computational demands. 

\subsection{Free corrosion potential} \label{ss:fcp}
Firstly, we discuss the calculation method for free corrosion potential, which is defined as the equilibrium potential between anodic and cathodic reactions. 
At this potential, the reaction rate for the anodic reaction is equal to that for the cathodic reaction. 
Phenomenologically, the Butler--Volmer equation describes the relationship between such currents and electrode potentials in the CT rate-determining step. 
However, the Butler--Volmer equation requires the simultaneous consideration of the anodic and cathodic reactions. 
To determine the free corrosion potential, the Tafel equation is useful, which is an approximated formula of the Butler--Volmer equation within the high voltage regime. 
The Tafel equation for the anodic and cathodic reactions with a single transferred electron is written as:
\begin{eqnarray}
i^{\mathrm{a(c)}} &=& i^{\rm a(c)}_{\mathrm{0}} \exp{\frac{\pm \alpha^{\mathrm{a(c)}}F \eta^\mathrm{a(c)}}{RT}},
\end{eqnarray}
where $i^{\rm a(c)}$ is the current for the anodic (cathodic) reaction, and $i^{\rm a(c)}_{\mathrm{0}}$ is referred to as the exchange current density. 
$\alpha^{\mathrm{a(c)}}$, $R$, and $\eta^\mathrm{a(c)}$ are the transfer coefficient, gas constant, and over-potential ($\eta^\mathrm{a(c)}=\Phi-E^\mathrm{a(c)}_{\mathrm{eq}}$), respectively. 
Using a logarithmic plot of $|i^{\rm a(c)}|$, the free corrosion potential is determined by the potential at the crossing point of the anodic and cathodic currents. The $\log{|i^{\rm a(c)}|}$--voltage plot is called the Tafel plot. This approach to determine the free corrosion potential is referred to as the Tafel extrapolation method. 
In the Tafel (or Butler--Volmer) equation, the exchange currents and the transfer coefficients can be determined by activation barriers and free energy profiles  for the anodic and cathodic reactions, respectively \cite{ikeshoji2017toward}. 
Instead of directly evaluating these parameters, in this study, we use the following simple model to determine the free corrosion potentials for studying all corrosive reactions: $\alpha^{\mathrm{a}}=\alpha^{\mathrm{c}}=0.5$ as used in the experiments discussed in Ref.~\cite{ma2017first}, and $|i^{\rm a}_{\mathrm{0}}|=|i^{\rm c}_{\mathrm{0}}|=i_{\mathrm{0}}$, which assumes that the anodic and cathodic reaction rate constants equal to each other. 
Based on these assumptions, the dependence of the Tafel equation is written as the difference in grand potentials between the left- and right-hand components of the reaction ($\Delta \Omega^{\rm a(c)}$) as follows:
\begin{eqnarray}
|i^{\rm a(c)}| &=& i_{\mathrm{0}} \exp(\alpha^{\mathrm{a(c)}} \beta \Delta \Omega^{\rm a(c)}), \label{curr} \\
\Delta \Omega^{\rm a(c)} &=& \Omega^{\rm a(c)}_{\rm R} - \Omega^{\rm a(c)}_{\rm L}, \label{exc}
\end{eqnarray}
where subscripts L and R denote the left- and right-hand components of the reactions, respectively. 
$\beta$ indicates the temperature factor. 
From the logarithmic plot of $|i^{\rm a(c)}|$ as a function of $\mu_{\rm e}$, we obtain the Tafel plot. 
As $\log{|i^{\rm a(c)}|}$ directly depends on $\Delta \Omega^{\rm a(c)}$, we obtain the free corrosion potential from $\mu_{\rm e}$ at the crossing point of $\Delta \Omega^{\rm a}$ and $\Delta \Omega^{\rm c}$. 

Next, we derive a more flexible formula for $\Delta \Omega^{\rm a(c)}$, 
which directly corresponds to the logarithmic plot of the anodic (cathodic) current, using the formulations of the thermodynamic method. 
For simplicity, we use the anodic reaction~(\ref{a3}) as a representative of the reformulation. 
Substituting Eqs.~(\ref{oal}) and (\ref{oar}) to Eq.~(\ref{exc}), we obtain: 
\begin{eqnarray}
\Delta \Omega^{\mathrm{a}} 
&=&1/3\Omega( \mathrm{Al}_{m-1}(\mathrm{s}) ) \nonumber \\
&+& 1/3 A( \mathrm{Al}^{3+}(\mathrm{sol}) ) - \mu_{\mathrm{e}} \nonumber \\
&-& 1/3\Omega( \mathrm{Al}_m(\mathrm{s})     ). \label{do1}
\end{eqnarray}
Here, we assume that the curvature of the ground potential (e.g., the interfacial capacitance and its potential change) 
and the PZC, which is the apex of the ground potential, remain unchanged before and after the reaction. 
Furthermore, $\Delta \Omega^{\mathrm{a}}$ shows the difference in the Helmholtz free energy before and after the reaction. 
Therefore, Eq.~(\ref{do1}) becomes:
\begin{eqnarray}
\Delta \Omega^{\mathrm{a}}&=&1/3 A(\mathrm{Al}_{m-1}(\mathrm{s})) \nonumber \\ 
&+& 1/3 A( \mathrm{Al}^{3+}(\mathrm{sol}) ) -\mu_{\mathrm{e}} \nonumber \\ 
&-& 1/3 A(\mathrm{Al}_{m}(\mathrm{s})) \nonumber \\
&=&G^{\mathrm{a}}_{\mathrm{R}} - G^{\mathrm{a}}_{\mathrm{L}} -\mu_{\mathrm{e}} \nonumber \\
&=&-FE^{\mathrm{a}}_{\rm eq} \ [\mathrm{vs.} \ \Phi_{\mathrm{S}}] -\mu_{\mathrm{e}}, \label{doa}
\end{eqnarray}
Here, we substitute Eqs.~(\ref{gal}) and (\ref{gar}) in the second equality,
and the EMF formulation [Eqs.~(\ref{thermo}) and (\ref{dG})] is used in the last equality. 
From the same procedure, the $\Delta \Omega^{\mathrm{c}}$ is given as:
\begin{eqnarray}
\Delta \Omega^{\mathrm{c}} = FE^{\mathrm{c}}_{\rm eq} \ [\mathrm{vs.} \ \Phi_{\mathrm{S}}] + \mu_{\mathrm{e}}. \label{doc}
\end{eqnarray}
Using Eqs.~(\ref{doa}) and (\ref{doc}), we analytically obtain the chemical potential at the crossing point of $\Delta \Omega^{\mathrm{a}}$ and $\Delta \Omega^{\mathrm{c}}$ under the present model. Thus, the Tafel extrapolation is analytically conducted as follows: 
\begin{eqnarray}
\! \! \! \! \!  \mu_{\mathrm{corr}}[\mathrm{vs.}\ \Phi_{\mathrm{S}}] \! \! \! \! \! &=& \nonumber -F \\
\! \! \! \! \! &\times& \! \! \! \! \! (E^{\mathrm{a}}_{\mathrm{eq}}[\mathrm{vs.}\ \Phi_{\mathrm{S}}] +E^{\mathrm{c}}_{\mathrm{eq}}[\mathrm{vs.}\ \Phi_{\mathrm{S}}]) / 2. \label{muthermo} 
\end{eqnarray}
Here, $\mu_{\mathrm{corr}}$ is the electron chemical potential for the free corrosion potential. 
Through this relation, the free corrosion potential $E_{\mathrm{corr}}$ is obtained as $-\mu_{\mathrm{corr}}/e$. 
Eq.~(\ref{muthermo}) is also flexible, similar to the thermodynamic method, because we obtain the free corrosion potential without grand potential calculations. 

\subsection{Computational details}
\begin{figure} [htb]
\centering\includegraphics[width=80mm]{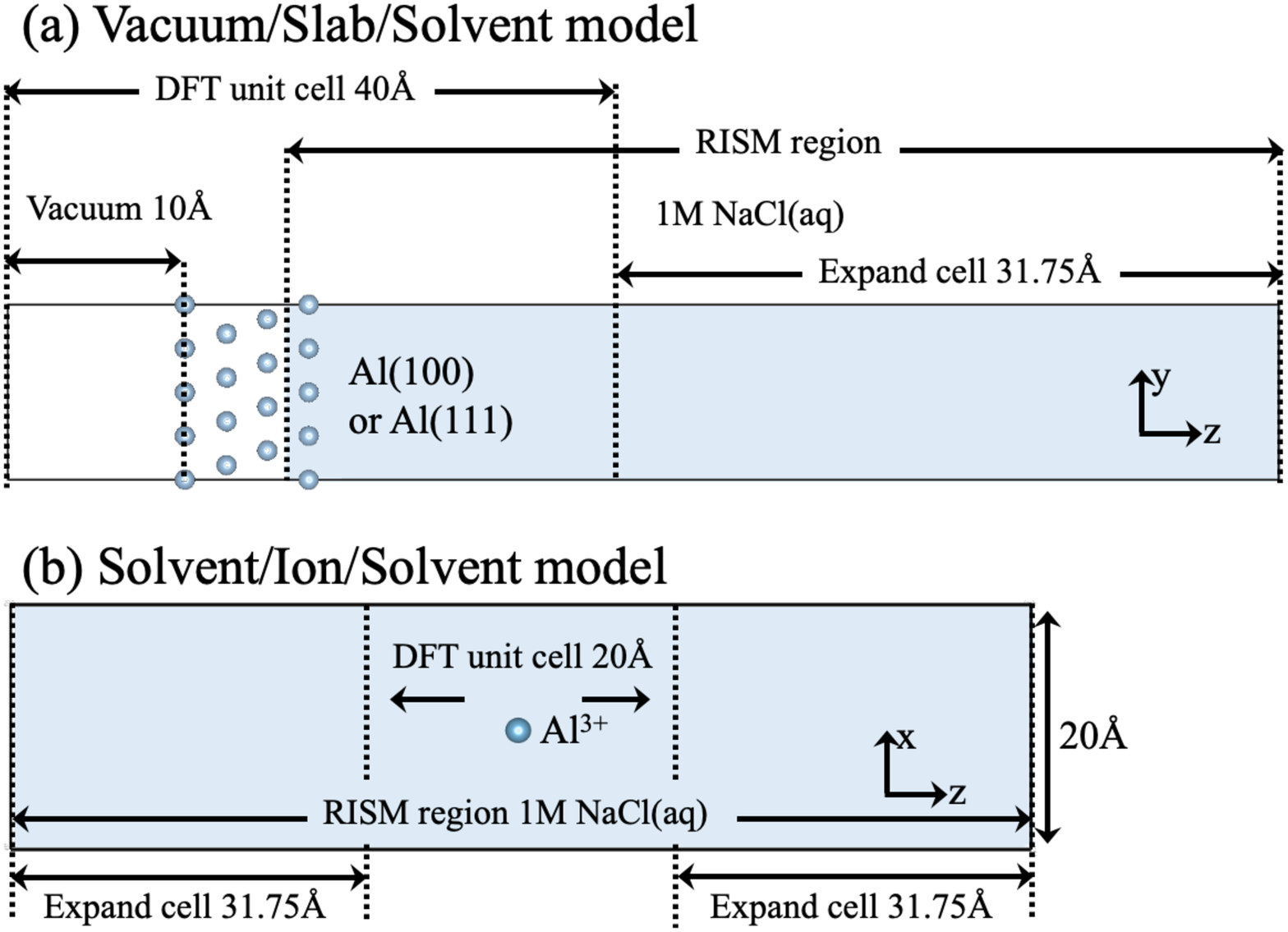}
\caption{ \label{cell}
(Color online). Schematic image of the simulation cells with boundary conditions of (a) vacuum/slab/solvent and (b) solvent/ion/solvent. Here, as a representative, only calculation cells for Al(111) and Al$^{3+}$ are shown. Cyan balls denote Al atom.}
\end{figure}

\begin{table}
\begin{center}
\caption{\label{lj}12-6 type of Lennard-Jones (LJ) parameters ($\varepsilon$ and $\sigma$) and charges ($Q$) for each site of RISM used in the present study are listed. 
Here, the solvent and solute denote components of the RISM and explicit particles, respectively. 
An implicit H$_2$O molecule is represented by the TIP5P model\cite{mahoney2000five} for precisely determining the hydration free energy for a H$_3$O$^+$ ion\cite{haruyama2018electrode}. 
The LJ parameter of Al$^{3+}$ was used to calculate the Al$^{3+}$(sol), AlCl$_3$(sol), AlCl$_4^{-}$(sol), and Al$_2$Cl$_6$(sol). }
\begin{tabular}{rrrr}
\hline
Atom & $\epsilon$[kcal/mol] & $\sigma$ [\AA] & $Q$[e] \\
\hline  \hline
Solvent & & & \\
H(H$_2$O)\cite{haruyama2018electrode} & 0.046 & 1.0 & +0.241 \\
O(H$_2$O)\cite{mahoney2000five} & 0.16 & 3.12 & 0.00 \\
L(H$_2$O)\cite{haruyama2018electrode} & 0.046 & 1.8 & -0.241 \\
H(H$_3$O$^+$)\cite{chuev2006quasilinear} & 0.046 & 0.4 & +0.5189 \\
O(H$_3$O$^+$)\cite{chuev2006quasilinear} & 0.1554 & 3.166 & -0.5567 \\
Na$^+$\cite{jorgensen2004free} & 0.002772 & 3.35 & +1.0 \\
Cl$^-$\cite{mayo1990dreiding} & 0.2833 & 3.9503 & -1.0 \\
Solute & & & \\
H\cite{haruyama2018electrode} & 0.046 & 1.0& -- \\
O\cite{hirschfelder2009intermolecular} & 0.1554 & 3.166 & -- \\
Al$^{3+}$ & 0.2120 & 1.850 & -- \\
Cl$^-$\cite{mayo1990dreiding} & 0.2833 & 3.9503 & -- \\
Al (slab)\cite{rappe1992uff} & 0.505 & 4.01 & -- \\
\hline
\end{tabular}
\end{center}
\end{table}

All calculations were carried out using plane-wave basis sets and ultrasoft-pseudopotential\cite{laasonen1991implementation, laasonen1993car, garrity2014pseudopotentials} code \textsc{Quantum Espresso}\cite{giannozzi2009quantum, giannozzi2017advanced}, which utilizes the ESM-RISM method \cite{nishihara2017hybrid}. 

For electronic structure calculations, the cut-off energies for wave functions and augmented charges were 40Ry and 320Ry, respectively. 
The spin-unpolarized version of the generalized gradient approximation parameterized by Perdew, Burke, and Ernzerhof\cite{perdew1996generalized} was used for the electron–electron exchange-correlation function. 
Furthermore, $\bf k$-point samplings of 16$\times$16$\times$16, 4$\times$4$\times$1, and only the $\Gamma$-point was used to calculate the Al bulk, Al slabs, and ions (or molecules), respectively. The occupation number of electrons was determined by the Gaussian smearing method with a smearing width of 0.01Ry. 
Four-layered slab models were used to represent Al(100) and Al(111) surfaces. 
We carried out cell optimization for the Al bulk and obtained an optimized cell parameter of 4.038\AA, which was in good agreement with the experimental lattice constant of 4.05\AA \cite{popovic1992lattice} and used to construct the surface slabs. 
We will further discuss the details of slab geometries and calculation cells later. 
For ions and molecules, we carried out structural optimization until all forces acting on the atoms became less than 1.0$\times$10$^{-3}$Ry/Bohr. 
To calculate the potentials, the $E_{\rm ZP}$ for bonded molecules and ions are determined using the \textsc{Gaussian16} program package \cite{g16} with BLYP/6-311+$G(3df, 2pd)$; the results are listed in Table.~\ref{zero}. For the other parameters, we used default values of \textsc{Gaussian16}.

\begin{table}[bth]
\begin{center}
\caption{\label{zero} Calculated zero-point vibration energy $E_{\rm ZP}$ unit in eV using \textsc{Gaussian16} program package \cite{g16} with condition of BLYP/6-311+$G(3df, 2pd)$. 
}
\begin{tabular}{lr}
\hline
& $E_{\rm ZP}$ [eV] \\
\hline  \hline
Cathode reaction& \\
H$_2$ & 0.270 \\
H$_2$O & 0.562 \\
H$_3$O$^+$ & 0.909 \\
Anode reaction & \\ 
AlCl$_3$ & 0.126 \\
AlCl$_4^-$ & 0.151 \\
Al$_2$Cl$_6$ & 0.284 \\
\hline
\end{tabular}
\end{center}
\end{table}

To evaluate the solute, the NaCl(aq) solution at a temperature of 300K was represented by the RISM \cite{hirata1981extended, hirata1982application, kovalenko1998three}. 
In the present study, 1M NaCl(aq) solutions under the conditions of pH=1, 2, and 7 were adopted. 
As pH=7 represents the neutral condition of the solution, it is out of the range of the present study. However, as a reference, we carried out calculations for neutral solutions.  
The concentrations of H$_2$O solvent and NaCl salt were 1.0cm$^3$/g (55.6mol/L) and 1mol/L, respectively. For calculations at pH=1 (or 2), 0.1 (or 0.01)mol/L HCl was added to the NaCl(aq). The calculation of the 1mol/L NaCl(aq) solution under the condition of pH=7 [NaCl(aq, pH=7)] was carried out without HCl. 
To solve the RISM equation, we used the Kovalenko and Hirata type of closure function\cite{kovalenko1998three}, and the convergence criterion of residuals of correlation functions in the RISM equation was 1$\times$10$^{-6}$. The cut-off energy for the reciprocal representation of the Laue-represented RISM equations was set to 160Ry. 
For the calculation of $\Delta \mu_{\rm solv}$, the Gaussian fluctuation method was used \cite{levy1991gaussian}. 
In the present study, we used the LJ type classical force field to describe site–site interactions. Table~\ref{lj} shows the LJ parameters and charges used for the RISM calculations. 
We note that the LJ parameter for the explicit Al$^{3+}$ ion was adjustably determined to reproduce the standard electrode potential for the reaction of Al(s)$\rightarrow$Al$^{3+}$(sol)+3e$^-$. 
Thus, the hydration effect on the explicit Al$^{3+}$ ion is well reproduced within the ESM-RISM theory.
The Lorentz–Berthelot combination rule\cite{lorentz1881ueber} was used for the LJ parameters for different types of atoms. 

For the ESM-RISM calculations, we used two boundary conditions: vacuum/slab/solvent (VSS) and solvent/ion/solvent (SIS). 
Figures~\ref{cell}(a) and (b) show the schematics of unit cells for the ESM-RISM calculation with each boundary condition. 
For surface calculations, Al(100) and Al(111) slabs had surface areas of 12.114$\times$12.114$\rm \AA^2$ and 8.565$\times$9.891\AA$^2$, respectively, with a length of 40$\rm \AA$ in the $z$-direction for the DFT cell. 
We carried out calculations of ions and molecules using a cubic cell with dimensions of 20$\times$20$\times$20\AA$^3$. 
In the RISM calculations, to obtain the converged distribution function of the solution, an expanded cell with a length of 31.75$\rm \AA$ in the $z$-direction was attached to the right-hand- and both sides of the calculation cells for the VSS and SIS, respectively. For VSS, we take a 10$\rm \AA$ of the vacuum region from the left-hand boundary of the calculation cell, whereby the ESM boundary condition of the vacuum is attached to the left-hand side of the unit cell.

\section{Results} \label{s:result}
In this section, we present the results obtained through the ESM-RISM method. 
Firstly, we describe the $\Omega$ profiles at the Al(100) and Al(111)/NaCl(aq, pH=1) interfaces with and without V$_{\mathrm{Al}}$. 
Secondly, we discuss the electrode potential results for the anodic and cathodic reactions, and then present the results of the Tafel plots for the anodic and cathodic reactions. 
Thirdly, we compare results of free corrosion potential between the electrochemical and thermodynamic methods. 
Thereafter, we appraise the pH-dependent behavior of the free corrosion potential relative to the previous experiment. 
Finally, the surface dependence of the free corrosion potentials and defect formation energies are presented. 

\subsection{Tafel plot from the electrochemical method}

\begin{figure} [htb]
\centering\includegraphics[width=75mm]{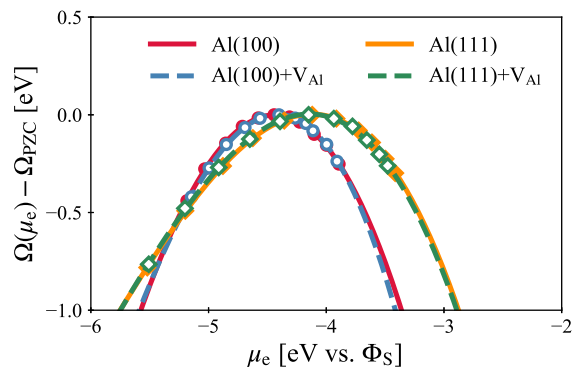}
\caption{ \label{grandfit}
(Color online). Grand potential profiles ($\Omega$) for Al(100)/ and Al(111)/1M NaCl(aq, pH=1) interfaces. Symbols denote raw data obtained by ESM-RISM calculations, and curves are the results of curve fitting using a polynomial function.}
\end{figure}

The results of the grand potential profile evaluation as a function of $\mu_{\rm e}$[vs.~$\Phi_{\rm S}$] for the Al(100) and Al(111)/1M NaCl(aq, pH=1) interfaces with and without V$_\mathrm{Al}$ are shown in Fig.~\ref{grandfit}. 
The $\Omega$ profiles are obtained by introducing excess charge ($\Delta N_{\rm e}$) to the electrode surface, where $\Delta N_{\rm e}$ changes from -1.0$e$/cell to +1.0$e$/cell at 0.2$e$/cell intervals. 
Here, we neglect the specific adsorption of halogen ions (Cl$^-$) on the surfaces. 
To compare $\Omega$ profiles between the surfaces, we select the maximum values of $\Omega$ ($\Omega_{\mathrm{PZC}}$) as the zero energy for each $\Omega$. 
As expected, all the $\Omega$ profile results show the approximate inverse of the parabola-shape centered at the point $\Delta N_{\rm e}=0$, which is consistent with the constant capacitance model. 
The $\mu_{\rm e}$[vs.~$\Phi_{\rm S}$] at $\Delta N_{\rm e}=0$ for clean Al(100) and Al(111)/1M NaCl(aq, pH=1) interfaces are $-4.44$eV and $-4.10$eV, respectively. 
Thus, the PZC for electrodes depends on the surface orientation, which is consistent with the findings of a previous experiment \cite{hamelin1982crystallographic}.
For defected surfaces, the curvatures and PZC of $\Omega$ are similar to those for clean surfaces. 
This result indicates that the change in the details of the EDL by introducing V$_{\mathrm{Al}}$ is relatively small. 
Thus, we expect that the difference in the results of equilibrium potentials between Eqs.~(\ref{ec}) and (\ref{thermo}) is rather small. 

Next, we fit the results of $\Omega$ by a third-order polynomial function, which is shown as solid lines in Fig.~\ref{grandfit}. 
The fitted curves duly described the original data obtained by the ESM-RISM calculations. 
We calculated $\Omega$ for all combinations of interfaces between Al surfaces and solutions and performed polynomial fitting for the results of $\Omega$. To simplify further analysis, hereafter, we will use the $\Omega$ profiles obtained by the polynomial fitting.

\begin{figure} [htb]
\centering\includegraphics[width=70mm]{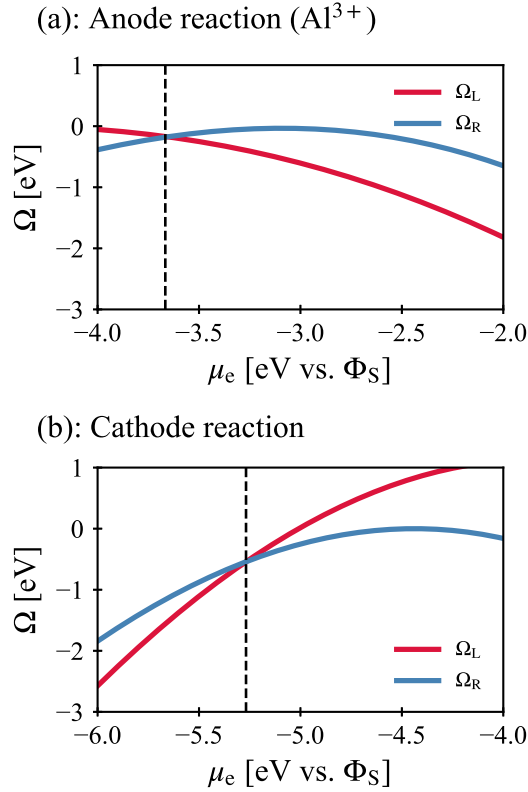}
\caption{ \label{electrode}
(Color online). Grand potential profiles ($\Omega$) of the left- and right-hand components for (a) anodic reaction with a product of Al$^{3+}$ and (b) cathodic reaction at Al(100)/1M NaCl(aq, pH=1) interfaces. The red and blue lines denote the $\Omega$ profile for the left- and right-hand components of each reaction ($\Omega_{\rm L}$ and $\Omega_{\rm R}$), respectively. $\Omega=0$ indicates $\Omega(\mu_{\mathrm{e}})-\Omega_{\mathrm{PZC}}$. As the anodic and cathodic reactions are CT reactions, $\Omega_{\mathrm{PZC}}$ denote the left- and right-hand components of the anodic and cathodic reactions, respectively. The vertical dashed lines indicate the equilibrium electron chemical potential for the corresponding reactions.}
\end{figure}

Figure~\ref{electrode} shows the results of the $\Omega$ profiles at the Al(100)/1M NaCl(aq, pH=1) interface for left- and right-hand components of anodic and cathodic reactions. 
Here, we only show the results of $\Omega$ using reaction~(\ref{a3}) as a representative of the anodic reactions. 
The equilibrium potentials for the anodic (cathodic) reaction were obtained by the chemical potentials at the crossing points of $\Omega^{\mathrm{a(c)}}_{\mathrm{L}}$ and $\Omega^{\mathrm{a(c)}}_{\mathrm{R}}$ ($\mu^{\mathrm{a(c)}}_{\mathrm{eq}}$). 
From the $\Omega$ profiles, the anodic reaction is reduced to a lower $\mu_{\mathrm{e}}$ than $\mu^{\mathrm{a}}_{\mathrm{eq}}$ because in such a reaction, $\Omega_{\mathrm{R}}$ becomes lower than $\Omega_{\mathrm{L}}$. 
Conversely, the cathodic reaction occurs at a higher $\mu_{\mathrm{e}}$ than $\mu^{\mathrm{a(c)}}_{\mathrm{eq}}$. 
Thus, the corrosion reaction occurs in the range of $\mu_{\mathrm{e}}$ from $\mu^{\mathrm{c}}_{\mathrm{eq}}$ to $\mu^{\mathrm{a}}_{\mathrm{eq}}$. 

The result of $\mu^{\mathrm{c}}_{\mathrm{eq}}$ corresponds to the $\mu_{\mathrm{e}}$ of the reversible hydrogen electrode (RHE) potential at pH=1, when it is measured from the inner potential. 
The electrode potential measured from the inner potential ($\Phi$[vs.~$\Phi_{\rm S}$]) was converted to that measured from RHE ($E$[vs.~RHE]) by the RHE potential relative to the inner potential ($E_{\rm RHE}$[vs.~$\Phi_{\rm S}]$) as follows:
\begin{eqnarray}
\! \! \! \! \! \! \! \! \! E [\mathrm{vs. \ RHE}] \! \! \! \! \! &=& \! \! \! \! \! \Phi [\mathrm{vs.} \ \Phi_{\rm S}] - E_{\rm RHE} [\mathrm{vs.} \ \Phi_{\rm S}] \nonumber \\
&=&\! \! \! \! \! -(\mu_{\rm e}[\mathrm{vs.} \ \Phi_{\rm S}] - \mu_{\rm RHE}[\mathrm{vs.} \ \Phi_{\rm S}])/e.
\end{eqnarray}
Here, $\mu_{\rm RHE}$[vs.~$\Phi_{\rm S}$] denotes $\mu_{\mathrm{eq}}$[vs.~$\Phi_{\rm S}$] corresponding to the RHE. 
Notably, $ \mu_{\rm RHE}$[vs.~$\Phi_{\rm S}$] is equal to the equilibrium potential of the cathodic reaction with respect to the inner potential of various solutions. 
In addition, the RHE potential ($E_{\rm RHE}$) measured from the inner potential at temperature $T$ can be represented using the SHE potential $E_{\rm SHE}$ as follows:
\begin{eqnarray}
\! \! \! \! \! E_{\rm RHE} [\mathrm{vs.} \ \Phi_{\rm S}] &=& E_{\rm SHE}[\mathrm{vs.} \ \Phi_{\rm S}] \nonumber \\
&-& \frac{k_{\rm B}T}{e}\ln(10)\times \mathrm{pH}. \label{rhe}
\end{eqnarray}
Here, $k_{\rm B}$ is the Boltzmann constant, and the coefficient of the second term in Eq.~(\ref{rhe}) is 0.059V at a temperature of 298K. In the ESM-RISM calculation, we can determine only $E_{\rm RHE}$[vs.~$\Phi_{\rm S}$] as a relative potential because we usually choose the inner potential as the reference potential level. Consequently, we cannot directly compare the electrode potentials using different solutions without the contact potential difference\cite{haruyama2018electrode}. However, we can convert the RHE potential to the SHE potential using Eq.~(\ref{rhe}). 
This conversion allows us to compare electrode potentials with different solutions obtained by the ESM-RISM, using the unique reference potential. 

\begin{figure} [htb]
\centering\includegraphics[width=70mm]{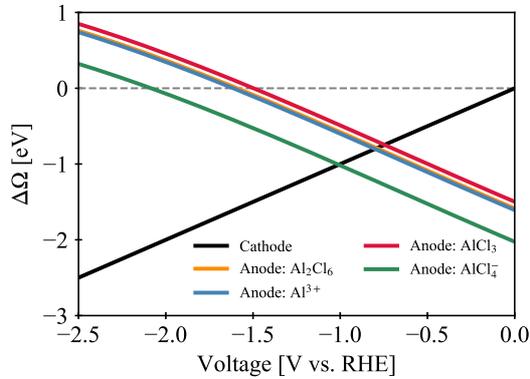}
\caption{ \label{tafel}
(Color online).~$\Delta \Omega$ as a function of voltage[V vs.~RHE] for the anodic and cathodic reactions at the Al(100)/1M NaCl(aq, pH=1) interface. 
The solid black line denotes the $\Delta \Omega$ for the cathodic reaction, while the blue, red, orange, and green lines indicate the $\Delta \Omega$ for the anodic reaction with a product of Al$^{3+}$, AlCl$_3$, Al$_2$Cl$_6$, and AlCl$_4^-$, respectively. The horizontal dashed line denotes $\Delta \Omega =0$.}
\end{figure}

Figure~\ref{tafel} shows the results of the logarithmic plot of $i^{\rm a(c)}$ at the Al(100)/1M NaCl(aq, pH=1) interface as a function of voltage vs.~RHE obtained by the ESM-RISM. 
Here, for simplicity, we only plot the $\Delta \Omega$, because $\beta$, $i_0$ and $\alpha^{\mathrm{a}(c)}$ are the constant and same values in Eq.~(\ref{curr}). 
The voltages at the crossing points of $\Delta \Omega$ for the cathodic and anodic reactions denote the free corrosion potentials for each anodic reaction, which are measured from the RHE potential. 
In addition, the voltages at $\Delta \Omega=0$ correspond to the equilibrium potentials measured from the RHE potential for each reaction. 
Thus, the anodic and cathodic reactions do not occur for voltages in the $\Delta \Omega>0$ range. 
From this plot for various pH conditions of NaCl(aq), we can obtain the pH dependence of the free corrosion potentials measured from the RHE potential, and thereafter calibrate the reference electrode potential from the RHE to the SHE using Eq.~(\ref{rhe}). 
 
\subsection{Free corrosion potentials from the electrochemical and thermodynamic methods}
\begin{table}[bth]
\begin{center}
\caption{\label{OvsG} Comparison between the results of free corrosion potentials by electrochemical ($E_{\mathrm{corr}}(\Omega)$) and thermodynamic ($\Delta G$) methods (unit in V vs.~RHE), as well as products of anodic reactions. Differences in the free corrosion potentials between the methods are also listed (unit in V vs.~RHE). All the results of the free corrosion potentials are calculated for the Al(100)/1M NaCl(aq, pH=1) interfaces. 
}
\begin{tabular}{lrrr}
\hline
Products & $E_{\mathrm{corr}}(\Omega)$ & $E_{\mathrm{corr}}(G)$ & Difference \\
\hline \hline
$\mathrm{Al^{3+}}$ & -0.819 & -0.810  & -0.008 \\
$\mathrm{AlCl_3}$ & -0.763 & -0.796 & 0.033 \\
$\mathrm{Al_2Cl_6}$ & -0.808 & -0.840 & 0.032 \\
$\mathrm{AlCl_4^-}$ & -1.029 & -1.068 & 0.040 \\
\hline
\end{tabular}
\end{center}
\end{table}
Here, we compare the results of the free corrosion potentials by the electrochemical and thermodynamic methods, as shown in Table~\ref{OvsG} with the products of the anodic reactions. 
When the change in the interfacial capacitance is negligibly small, the free corrosion potentials obtained by both methods have the same value. 
We found that the results of free corrosion potentials by the electrochemical method ($E_{\mathrm{corr}}(\Omega)$) were in reasonable agreement with those obtained using the thermodynamic method ($E_{\mathrm{corr}}(G)$). 
The differences between $E_{\mathrm{corr}}(\Omega)$ and $E_{\mathrm{corr}}(G)$ originate from the deviation of $\Delta \Omega$ from the linear function. 
This difference in $\Delta \Omega$ is regarded as the change in the interfacial capacitance due to the formation of V$_{\mathrm{Al}}$. 
Thus, the change in the details of EDL slightly affects the free corrosion potential (the absolute value of the difference in the free corrosion potential between the methods is an order of error within 0.01V).

\subsection{pH dependence of free corrosion potentials in acidic region}
 
\begin{figure} [htb]
\centering\includegraphics[width=70mm]{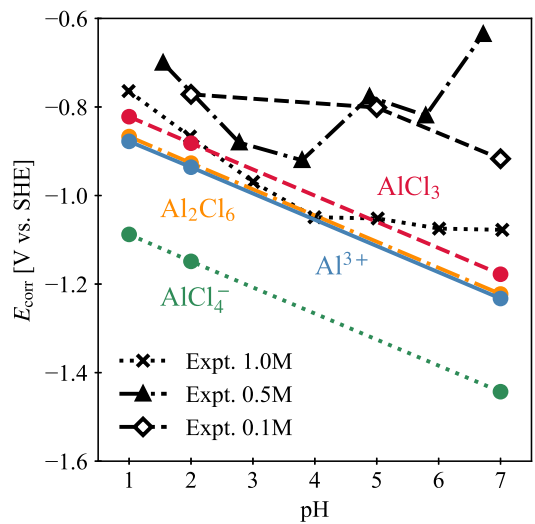}
\caption{ \label{ecorr}
(Color online). Free corrosion potential as a function of pH (unit in V vs.~SHE). Black crosses, triangles, and diamonds, respectively, denote the experimental data of free corrosion potentials for interfaces between the Al electrode and the 1M, 0.5M, and 0.1M NaCl(aq) solution obtained from Refs.~\citenum{minakawa19841},~\citenum{dibari1971electrochemical}, and~\citenum{mansouri2011anodic}. The red, blue, and green circles show the free corrosion potentials for AlCl$_3$, AlCl$_4^{-}$, and Al$_2$Cl$_6$, respectively. The solid, dotted, dashed, and dash-dotted lines denote for guides of the reader's eye. }
\end{figure}

Figure~\ref{ecorr} presents the results of free corrosion potential vs.~SHE for various conditions of the NaCl(aq) solution, as well as the experimental data for the 1M, 0.5M, and 0.1M NaCl(aq) solutions, which are, respectively, obtained from Refs.~\citenum{minakawa19841},~\citenum{dibari1971electrochemical} and~\citenum{mansouri2011anodic}. 
The free corrosion potentials for Al$^{3+}$, AlCl$_3$, and Al$_2$Cl$_6$ reasonably reproduce the experimental results. 
Conversely, the results of $E_{\rm corr}$ for AlCl$_4^-$ are lower than those for other products. 
This result suggests that the main corrosion products of the anodic reaction due to H$_2$ creation are Al$^{3+}$, AlCl$_3$, and Al$_2$Cl$_6$. 
The calculated $E_{\rm corr}$ decreases with increasing pH. 
This tendency also reproduce the experimental data under acidic conditions. 
In the present study, the pH dependence of $E_{\rm corr}$ comes from the second term of the Nernst's equation represented by Eq.~(\ref{rhe}). 
The potential of the H$_2$ creation reaction decreases with the increase in the pH. 
As a result of this, $E_{\rm corr}$ located between the potentials for the anode and cathode reactions decreases with increasing the pH. 
Therefore, the present model of the corrosive reaction duly reproduces the experimental data at the Al electrode in the acidic NaCl(aq) solution. 
Although the results of $E_{\rm corr}$ for the Al/NaCl(aq, pH=7) interface are reference data, we compared their results to those of the experiment.
Compared to the acidic region, the calculated $E_{\rm corr}$ for the neutral condition of NaCl(aq) deviates from the experimental data. 
Thus, the present model is limited in the acidic condition of the NaCl(aq) solution. 

\subsection{Surface dependence of free corrosion potentials}

\begin{figure} [htb]
\centering\includegraphics[width=70mm]{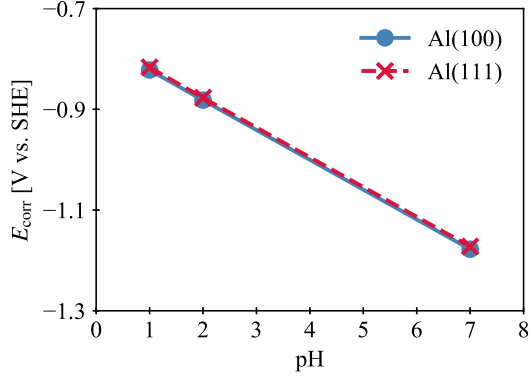}
\caption{ \label{ecorrs}
(Color online). Free corrosion potentials $E_{\mathrm{corr}}$ for AlCl$_3$ as a function of pH for different Al surfaces (unit in V vs.~SHE). 
The blue and red symbols denote the $E_{\mathrm{corr}}$ for Al(100) and Al(111), respectively. 
The solid and dashed lines, respectively, indicate guides for the reader's eye. 
}
\end{figure}

\begin{table}[bth]
\begin{center}
\caption{\label{vacancy} Vacancy formation energies for various interfaces. 
}
\begin{tabular}{lrr}
\hline 
Interface & Al(100) & Al(111) \\
\hline \hline
Vacuum & 0.54eV & 0.58eV \\
NaCl(aq, pH=1) & 0.60eV & 0.65eV \\ 
NaCl(aq, pH=2) & 0.61eV & 0.65eV \\ 
NaCl(aq, pH=7) & 0.61eV & 0.65eV \\ 
\hline
\end{tabular}
\end{center}
\end{table}

Here, we show the results of $E_{\mathrm{corr}}$ depending on the surface orientation, which are presented in Fig.~\ref{ecorrs}. 
We found that the results of $E_{\mathrm{corr}}$ for Al(100) are approximately the same as those for Al(111). 
The main contribution of the surface dependence of $E_{\mathrm{corr}}$ is attributed to the stability of the Al atoms on the outermost surfaces. 
Here, we show the results of the vacancy formation energy for the outermost Al atom ($E({\mathrm{V_{Al}}})$), calculated as:
\begin{eqnarray}
\! \! E({\mathrm{V_{Al}}})= \{ A(\mathrm{Al}_{m-1})&+&E_{\mathrm{DFT}}(\mathrm{Al}) \} \nonumber \\
&-&A(\mathrm{Al}_m). 
\end{eqnarray}
The results of $E({\mathrm{V_{Al}}})$ are listed in table~\ref{vacancy}. 
The results of $E({\mathrm{V_{Al}}})$ for the vacuum/Al interfaces agree with those of previous DFT calculations \cite{neugebauer1992adsorbate, polatoglou1993vacancy, stumpf1996ab}. 
For all interfaces, $E({\mathrm{V_{Al}}})$ at Al(100) is slightly smaller than that at Al(111). 
Thus, the stability of the Al atoms, depending on the surface orientation, is weak. 
This explains the small difference in the results of $E_{\mathrm{corr}}$ between Al(100) and Al(111). 

\section{Discussions} \label{s:discussion}

In this section, we first discuss the comparison between the electrochemical and thermodynamic methods. 
Secondly, we discuss the kinetics for the corrosive reaction. 
Then, the pH dependence of the free corrosion potential and corrosive reactions is discussed. 
Finally, we elucidate the surface-dependent corrosion. 

If the changes in capacitance and the PZC potential before and after the reaction are negligible, the thermodynamic method is an effective way to reduce the computational cost.
This is because the calculation of the SIS is much cheaper than that of the VSS. 
Therefore, this method is useful for a systematic study of $E_{\rm corr}$ with varying parameters, such as pH and temperature. 
However, it is necessary to simulate the change in the EDL, which plays a central role, using electrochemical methods, such as studies on the adsorption of ions (or molecules) \cite{weitzner2020toward}, and the desorption/insertion processes of ions \cite{haruyama2018analysis}.

Generally, under the atmospheric condition, the Al surface is covered with the passivation layer of oxides from the viewpoint of the experiment. 
In Ref.~\citenum{minakawa19841}, to avoid the effect of the passivation layers on the $E_{\rm corr}$, they are removed by controlling the bias potential. Therefore, the present results of $E_{\rm corr}$ using clean Al surfaces are closer to the experimental data taken from Ref.~\citenum{minakawa19841}. 
From this, the ESM-RISM method satisfactorily reproduces the nature of the solid/liquid interface under acidic conditions. 
In contrast, the difference in the results for the neutral NaCl(aq) solution between the theoretical and experimental data originates from the formation of a passive layer. 
According to the previous study \cite{verma2017corrosion} and Pourbaix Diagram \cite{zhu2010corrosion}, the Al surface is covered with a passive layer of oxide or hydroxide under the conditions of pH$>$4. 
The passive layers ensue from the dissolved O$_2$ and OH$^-$ participating in the cathode reaction and the hydrate complex of Al$^{3+}$\cite{yasuda1990pitting}; thus, the passive layer alters the details of the corrosion reactions and the free corrosion potential. 
Therefore, the model of corrosion used in the present study does not reproduce the free corrosion potential for neutral NaCl(aq) solutions. 
To determine the whole behavior of the pH dependence of $E_{\rm corr}$, we should systematically consider other anodic and cathodic reactions. 
We must investigate such systematic considerations on this corrosion model in future studies. 

In this study, we assumed the values of the exchange current density and the charge transfer coefficient, and it remains an open question on theoretically how to determine these kinetic parameters. 
For more sophisticated simulations with the kinetic effect, it is necessary to calculate the activation energy profiles for the anodic and cathodic reactions at the Al/NaCl(aq) interface, respectively. 
In the previous study based on the first-principles calculation \cite{ma2017first}, the exchange current density for the anodic reaction was evaluated by incorporating terms for surface, defect formation, and halogen adsorption energies into the activation energy of metal ions dissolved from the bulk into the electrolyte. 
Besides, Haruyama et al. \cite{haruyama2018analysis} and Ikeshoji and Otani \cite{ikeshoji2017toward} directly studied the activation profiles for the ion-dissolution and oxygen reduction reactions at the solid/liquid interface, respectively. 
In Ref.~\citenum{haruyama2018analysis}, the ESM-RISM method was applied to the Li$^+$ ion insertion/dissolution reaction at the graphite/1M LiPF$_6$ EC interface. In Ref.~\citenum{ikeshoji2017toward}, the oxygen reduction reaction at the Pt/water interface was simulated from the first-principles. Both the simulations well reproduce the experimental activation energy and cyclic voltammetry, respectively. 
After determining the kinetic parameters for the anodic and cathodic reactions using the above methods, we can obtain the free corrosion potential without the assumptions for the exchange-current densities and transfer coefficients using the Tafel extrapolation method presented in this study. 
It is a future task to study the free corrosion potential beyond the current model. 

Finally, we briefly discuss the surface dependence of the corrosion potential. 
The present results of free corrosion potentials weakly depend on the surface orientations. 
However, a previous experiment for single-crystal Al shows that the order of the difference in the pitting corrosion potential between the surface orientations is approximately 10mV \cite{yasuda1990pitting}, which is higher than the present results of $E_{\rm corr}$. 
This difference between the present and previous studies is mainly due to the absence of Cl$^-$ adsorption in the present study. 
The previous experiment suggests that pitting corrosion occurs via halogen adsorption onto the Al surface \cite{yasuda1990pitting}. 
Thus, to further understand the surface dependence of the corrosion reaction, we need to consider the halogen adsorption on individual Al surfaces. 

\section{Summary} \label{s:summary}

We investigated the free corrosion potential at the Al/NaCl(aq) interface using the ESM-RISM method. 
Firstly, we considered a set of corrosion reactions and formulated a method to determine the free corrosion potentials using the ESM-RISM. 
The equilibrium potentials for the anodic and cathodic reactions were determined from the results of the grand potentials. 
Thereafter, we obtained the Tafel plot by evaluating the difference in the grand potentials between the right- and left-hand components of the reactions. 
The results of the free corrosion potentials between the electrochemical and thermodynamic methods was compared. 
The results of the free corrosion potentials between these two methods agreed with each other within the order of 0.01V. 
Therefore, we concluded that the change in the EDL before and after the reaction was inconsiderable. 
In such a case, we submit that the thermodynamic method is more suitable for obtaining the free corrosion potential owing to a lower calculation cost. 

The pH dependence of the obtained free corrosion potentials was compared to the experimental results with the reference potential of the SHE. 
The pH dependence of the free corrosion potentials under acidic conditions reasonably reproduce the previous experimental results. 
This result indicates that the ESM-RISM correctly describes the environmental effects of NaCl(aq) solutions. 
Nonetheless, the results for neutral NaCl(aq) solutions deviated from the obtained experimental data. This phenomenon suggests that we need to consider the formation of the passive layer on the electrode in the case of neutral NaCl(aq) solutions.

The following point should be noted here. As a future study, by computing the activation energy profile, we will be able to improve the treatment of the kinetic effect of the corrosion process, and calculate the free corrosion potential without making assumptions for the exchange current density and transfer coefficient. 

Finally, we examined the dependence of the free corrosion potential on the surface orientations. 
The difference in the free corrosion potentials between Al(100) and Al(111) is small. 
This result is consistent with the results of the vacancy formation energies. 
From these results, we conclude that the areal dependence of the free corrosion potential at the Al/NaCl(aq) interface is small. 
However, by considering the pitting corrosion, the corrosion potential depends on the surface orientations, which may be caused by Cl$^-$ adsorption. 
To better understand the concept of corrosion reactions, we additionally need to consider halogen adsorption on the surface.

\section*{Acknowledgments}
The authors thank Prof.~J\"{o}rg Neugebauer and Dr.~Mira Todorova for a fruitful discussion. 
The authors also thank Dr.~Tamio Ikeshoji for a variable discussion. 
S.H. and M.O. acknowledge the partial financial support from MEXT as 
“Program for Promoting Researches on the Supercomputer Fugaku” (Fugaku Battery \& Fuel Cell Project), Grant Number JPMXP1020200301. 
The computations were performed using the supercomputers of the Information Technology Center at the University of Tokyo; OCTOPUS was provided by Osaka University and Flow was provided by Nagoya University.

\bibliography{mybibfile}

\begin{thebibliography}{10}
\expandafter\ifx\csname url\endcsname\relax
  \def\url#1{\texttt{#1}}\fi
\expandafter\ifx\csname urlprefix\endcsname\relax\def\urlprefix{URL }\fi
\expandafter\ifx\csname href\endcsname\relax
  \def\href#1#2{#2} \def\path#1{#1}\fi

\bibitem{bethencourt1998lanthanide}
M.~Bethencourt, F.~Botana, J.~Calvino, M.~Marcos, M.~Rodriguez-Chacon,
  Lanthanide compounds as environmentally-friendly corrosion inhibitors of
  aluminium alloys: a review, Corros. Sci. 40~(11) (1998) 1803--1819.

\bibitem{kritzer2004corrosion}
P.~Kritzer, Corrosion in high-temperature and supercritical water and aqueous
  solutions: a review, J. Supercrit. Fluid. 29~(1-2) (2004) 1--29.

\bibitem{chen2011review}
X.~Chen, N.~Birbilis, T.~Abbott, Review of corrosion-resistant conversion
  coatings for magnesium and its alloys, Corr. 67~(3) (2011) 035005--1.

\bibitem{verma2017corrosion}
C.~Verma, E.~E. Ebenso, M.~Quraishi, Corrosion inhibitors for ferrous and
  non-ferrous metals and alloys in ionic sodium chloride solutions: A review,
  J. Mol. Liq. 248 (2017) 927--942.

\bibitem{dibari1971electrochemical}
G.~Dibari, H.~Read, Electrochemical behavior of high purity aluminum in
  chloride containing solutions, Corr. 27~(11) (1971) 483--494.

\bibitem{minakawa19841}
M.~Kazuyasu, N.~Michinori, S.~Eiichi, Anodic polarization behavior of high
  purity aluminum in acidic and neutral solutions, J. Jpn. Inst. Met. 34~(12)
  (1984) 702--707 (in japanese).

\bibitem{simoes2008assessment}
A.~Sim{\~o}es, D.~Battocchi, D.~Tallman, G.~Bierwagen, Assessment of the
  corrosion protection of aluminium substrates by a {M}g-rich primer: {EIS},
  {SVET} and {SECM} study, Prog. Org. Coat. 63~(3) (2008) 260--266.

\bibitem{mansouri2011anodic}
K.~Mansouri, K.~Ibrik, N.~Bensalah, A.~Abdel-Wahab, Anodic dissolution of pure
  {A}luminum during electrocoagulation process: {I}nfluence of supporting
  electrolyte, initial p{H}, and current density, Ind. Eng. Chem. Res. 50~(23)
  (2011) 13362--13372.

\bibitem{sherif2011effects}
E.-S.~M. Sherif, A.~Almajid, F.~H. Latif, H.~Junaedi, Effects of graphite on
  the corrosion behavior of aluminum-graphite composite in sodium chloride
  solutions, Int. J. Electrochem. Sci 6 (2011) 1085--1099.

\bibitem{acosta2014power}
G.~Acosta, L.~Veleva, J.~L{\'o}pez, Power spectral density analysis of the
  corrosion potential fluctuation of aluminium in early stages of exposure to
  {C}aribbean sea water, Int. J. Electrochem. Sci 9 (2014) 6464--6474.

\bibitem{mondal2016development}
J.~Mondal, A.~Marques, L.~Aarik, J.~Kozlova, A.~Sim{\~o}es, V.~Sammelselg,
  Development of a thin ceramic-graphene nanolaminate coating for corrosion
  protection of stainless steel, Corros. Sci. 105 (2016) 161--169.

\bibitem{ropo2007theoretical}
M.~Ropo, K.~Kokko, M.~Punkkinen, S.~Hogmark, J.~Koll{\'a}r, B.~Johansson,
  L.~Vitos, Theoretical evidence of the compositional threshold behavior of
  {F}e{C}r surfaces, Phys. Rev. B 76~(22) (2007) 220401.

\bibitem{taylor2008first}
C.~D. Taylor, M.~Neurock, J.~R. Scully, First-principles investigation of the
  fundamental corrosion properties of a model {C}u38 nanoparticle and the
  (111),(113) surfaces, J. Electrochem. Soc. 155~(8) (2008) C407.

\bibitem{ropo2011first}
M.~Ropo, K.~Kokko, E.~Airiskallio, M.~P.~J. Punkkinen, S.~Hogmark,
  J.~Koll{\'a}r, B.~Johansson, L.~Vitos, First-principles atomistic study of
  surfaces of {F}e-rich {F}e--{C}r, J. Phy.: Cond. Matter 23~(26) (2011)
  265004.

\bibitem{ma2017first}
H.~Ma, X.-Q. Chen, R.~Li, S.~Wang, J.~Dong, W.~Ke, First-principles modeling of
  anisotropic anodic dissolution of metals and alloys in corrosive
  environments, Acta Materialia 130 (2017) 137--146.

\bibitem{ke2019density}
H.~Ke, C.~D. Taylor, Density functional theory: an essential partner in the
  integrated computational materials engineering approach to corrosion, Corr.
  75~(7) (2019) 708--726.

\bibitem{otani2008structure}
M.~Otani, I.~Hamada, O.~Sugino, Y.~Morikawa, Y.~Okamoto, T.~Ikeshoji, Structure
  of the water/platinum interface--a first principles simulation under bias
  potential, Phys. Chem. Chem. Phys. 10~(25) (2008) 3609--3612.

\bibitem{ikeshoji2012charged}
T.~Ikeshoji, M.~Otani, I.~Hamada, O.~Sugino, Y.~Morikawa, Y.~Okamoto, Y.~Qian,
  I.~Yagi, The charged interface between {P}t and water: {F}irst principles
  molecular dynamics simulations, AIP Advances 2~(3) (2012) 032182.

\bibitem{ikeshoji2017toward}
T.~Ikeshoji, M.~Otani, Toward full simulation of the electrochemical oxygen
  reduction reaction on {P}t using first-principles and kinetic calculations,
  Phys. Chem. Chem. Phys. 19~(6) (2017) 4447--4453.

\bibitem{ambrosio2018ph}
F.~Ambrosio, J.~Wiktor, A.~Pasquarello, p{H}-{D}ependent {S}urface {C}hemistry
  from {F}irst {P}rinciples: {A}pplication to the {B}i{VO}$_4$ (010)--{W}ater
  {I}nterface, ACS Appl. Mater. Inter. 10~(12) (2018) 10011--10021.

\bibitem{wiktor2019electron}
J.~Wiktor, A.~Pasquarello, Electron and {H}ole {P}olarons at the
  {B}i{VO}$_4$--{W}ater {I}nterface, ACS Appl. Mater. Inter. 11~(20) (2019)
  18423--18426.

\bibitem{mathew2014implicit}
K.~Mathew, R.~Sundararaman, K.~Letchworth-Weaver, T.~Arias, R.~G. Hennig,
  Implicit solvation model for density-functional study of nanocrystal surfaces
  and reaction pathways, J. Chem. Phys. 140~(8) (2014) 084106.

\bibitem{sakong2015density}
S.~Sakong, M.~Naderian, K.~Mathew, R.~G. Hennig, A.~Gro{\ss}, Density
  functional theory study of the electrochemical interface between a {P}t
  electrode and an aqueous electrolyte using an implicit solvent method, J.
  Chem. Phys. 142~(23) (2015) 234107.

\bibitem{lespes2015using}
N.~Lespes, J.-S. Filhol, Using implicit solvent in ab initio electrochemical
  modeling: investigating {L}i$^+$/{L}i electrochemistry at a {L}i/{S}olvent
  interface, Journal of chemical theory and computation 11~(7) (2015)
  3375--3382.

\bibitem{sundararaman2017evaluating}
R.~Sundararaman, K.~Schwarz, Evaluating continuum solvation models for the
  electrode-electrolyte interface: {C}hallenges and strategies for improvement,
  J. Chem. Phys. 146~(8) (2017) 084111.

\bibitem{otani2006first}
M.~Otani, O.~Sugino, First-principles calculations of charged surfaces and
  interfaces: {A} plane-wave nonrepeated slab approach, Phys. Rev. B 73~(11)
  (2006) 115407.

\bibitem{hamada2013improved}
I.~Hamada, O.~Sugino, N.~Bonnet, M.~Otani, Improved modeling of electrified
  interfaces using the effective screening medium method, Phys. Rev. B 88~(15)
  (2013) 155427.

\bibitem{hirata1981extended}
F.~Hirata, P.~J. Rossky, An extended {RISM} equation for molecular polar
  fluids, Chem. Phys. Lett. 83~(2) (1981) 329--334.

\bibitem{hirata1982application}
F.~Hirata, B.~M. Pettitt, P.~J. Rossky, Application of an extended {RISM}
  equation to dipolar and quadrupolar fluids, J. Chem. Phys. 77~(1) (1982)
  509--520.

\bibitem{kovalenko1998three}
A.~Kovalenko, F.~Hirata, Three-dimensional density profiles of water in contact
  with a solute of arbitrary shape: a {RISM} approach, Chem. Phys. Lett.
  290~(1-3) (1998) 237--244.

\bibitem{nishihara2017hybrid}
S.~Nishihara, M.~Otani, Hybrid solvation models for bulk, interface, and
  membrane: {R}eference interaction site methods coupled with density
  functional theory, Phys. Rev. B 96~(11) (2017) 115429.

\bibitem{haruyama2018electrode}
J.~Haruyama, T.~Ikeshoji, M.~Otani, Electrode potential from density functional
  theory calculations combined with implicit solvation theory, Phys. Rev.
  Mater. 2~(9) (2018) 095801.

\bibitem{bonnet2012first}
N.~Bonnet, T.~Morishita, O.~Sugino, M.~Otani, First-{P}rinciples {M}olecular
  {D}ynamics at a {C}onstant {E}lectrode {P}otential, Phys. Rev. Lett. 109~(26)
  (2012) 266101.

\bibitem{weitzner2020toward}
S.~E. Weitzner, S.~A. Akhade, J.~B. Varley, B.~C. Wood, M.~Otani, S.~E. Baker,
  E.~B. Duoss, {T}oward {E}ngineering of {S}olution {M}icroenvironments for the
  {CO}$_2$ {R}eduction {R}eaction: {U}nraveling p{H} and {V}oltage {E}ffects
  from a {C}ombined {D}ensity-{F}unctional--{C}ontinuum {T}heory, J. Phys.
  Chem. Lett. 11~(10) (2020) 4113--4118.

\bibitem{yasuda1990pitting}
M.~Yasuda, F.~Weinberg, D.~Tromans, Pitting {C}orrosion of {A}l and {A}l-{C}u
  {S}ingle {C}rystals, J. Electrochem. Soc. 137~(12) (1990) 3708.

\bibitem{databese}
\url{http://www2.ucdsb.on.ca/tiss/stretton/database/inorganic_thermo.htm}.

\bibitem{johnson2016small}
J.~Johnson, D.~A. Case, T.~Yamazaki, S.~Gusarov, A.~Kovalenko, T.~Luchko, Small
  molecule hydration energy and entropy from 3{D}-{RISM}, J. Phys.: Condens.
  Matter 28~(34) (2016) 344002.

\bibitem{haruyama2018analysis}
J.~Haruyama, T.~Ikeshoji, M.~Otani, Analysis of lithium insertion/desorption
  reaction at interfaces between graphite electrodes and electrolyte solution
  using density functional+implicit solvation theory, J. Phys. Chem. C 122~(18)
  (2018) 9804--9810.

\bibitem{haruyama2019two}
J.~Haruyama, K.-i. Okazaki, Y.~Morita, H.~Nakamoto, E.~Matsubara, T.~Ikeshoji,
  M.~Otani, Two-{P}hase {R}eaction {M}echanism for {F}luorination and
  {D}efluorination in {F}luoride-{S}huttle {B}atteries: {A}
  {F}irst-{P}rinciples {S}tudy, ACS Appl. Mater. Inter. 12~(1) (2019) 428--435.

\bibitem{wang2006oxidation}
L.~Wang, T.~Maxisch, G.~Ceder, Oxidation energies of transition metal oxides
  within the {GGA+U} framework, Phys. Rev. B 73~(19) (2006) 195107.

\bibitem{grindy2013approaching}
S.~Grindy, B.~Meredig, S.~Kirklin, J.~E. Saal, C.~Wolverton, Approaching
  chemical accuracy with density functional calculations: {D}iatomic energy
  corrections, Phys. Rev. B 87~(7) (2013) 075150.

\bibitem{mahoney2000five}
M.~W. Mahoney, W.~L. Jorgensen, A five-site model for liquid water and the
  reproduction of the density anomaly by rigid, nonpolarizable potential
  functions, J. Chem. Phys. 112~(20) (2000) 8910--8922.

\bibitem{chuev2006quasilinear}
G.~Chuev, S.~Chiodo, S.~Erofeeva, M.~Fedorov, N.~Russo, E.~Sicilia, A
  quasilinear {RISM} approach for the computation of solvation free energy of
  ionic species, Chem. Phys. Lett. 418~(4-6) (2006) 485--489.

\bibitem{jorgensen2004free}
W.~L. Jorgensen, J.~P. Ulmschneider, J.~Tirado-Rives, Free energies of
  hydration from a generalized {B}orn model and an {ALL}-atom force field, J.
  Phys. Chem. B 108~(41) (2004) 16264--16270.

\bibitem{mayo1990dreiding}
S.~L. Mayo, B.~D. Olafson, W.~A. Goddard, {DREIDING}: a generic force field for
  molecular simulations, J. Phys. Chem. 94~(26) (1990) 8897--8909.

\bibitem{hirschfelder2009intermolecular}
J.~O. Hirschfelder, Intermolecular forces, Vol.~24, John Wiley \& Sons, 2009.

\bibitem{rappe1992uff}
A.~K. Rapp{\'e}, C.~J. Casewit, K.~Colwell, W.~A. Goddard~III, W.~M. Skiff,
  {UFF}, a full periodic table force field for molecular mechanics and
  molecular dynamics simulations, J. A. Chem. Soc. 114~(25) (1992)
  10024--10035.

\bibitem{laasonen1991implementation}
K.~Laasonen, R.~Car, C.~Lee, D.~Vanderbilt, Implementation of ultrasoft
  pseudopotentials in ab initio molecular dynamics, Phys. Rev. B 43~(8) (1991)
  6796.

\bibitem{laasonen1993car}
K.~Laasonen, A.~Pasquarello, R.~Car, C.~Lee, D.~Vanderbilt, Car-{P}arrinello
  molecular dynamics with {V}anderbilt ultrasoft pseudopotentials, Phys. Rev. B
  47~(16) (1993) 10142.

\bibitem{garrity2014pseudopotentials}
K.~F. Garrity, J.~W. Bennett, K.~M. Rabe, D.~Vanderbilt, Pseudopotentials for
  high-throughput {DFT} calculations, Comput. Mater. Sci. 81 (2014) 446--452.

\bibitem{giannozzi2009quantum}
P.~Giannozzi, S.~Baroni, N.~Bonini, M.~Calandra, R.~Car, C.~Cavazzoni,
  D.~Ceresoli, G.~L. Chiarotti, M.~Cococcioni, I.~Dabo, et~al., {QUANTUM
  ESPRESSO}: a modular and open-source software project for quantum simulations
  of materials, J. phys.: Cond. matter 21~(39) (2009) 395502.

\bibitem{giannozzi2017advanced}
P.~Giannozzi, O.~Andreussi, T.~Brumme, O.~Bunau, M.~B. Nardelli, M.~Calandra,
  R.~Car, C.~Cavazzoni, D.~Ceresoli, M.~Cococcioni, et~al., Advanced
  capabilities for materials modelling with {Q}uantum {ESPRESSO}, J. Phys.:
  Cond. Matter 29~(46) (2017) 465901.

\bibitem{perdew1996generalized}
J.~P. Perdew, K.~Burke, M.~Ernzerhof, Generalized gradient approximation made
  simple, Phys. Rev. Lett. 77~(18) (1996) 3865.

\bibitem{popovic1992lattice}
S.~Popovi{\'c}, B.~Gr{\v{z}}eta, V.~Ilakovac, R.~Kroggel, G.~Wendrock,
  H.~L{\"o}ffler, Lattice constant of the {F.C.C.} {A}l-rich $\alpha$-phase of
  {A}l-{Z}n alloys in equilibrium with {GP} zones and the $\beta$({Z}n)-phase,
  Phys. Status Solidi A 130~(2) (1992) 273--292.

\bibitem{g16}
M.~J. Frisch, G.~W. Trucks, H.~B. Schlegel, G.~E. Scuseria, M.~A. Robb, J.~R.
  Cheeseman, G.~Scalmani, V.~Barone, G.~A. Petersson, H.~Nakatsuji, X.~Li,
  M.~Caricato, A.~V. Marenich, J.~Bloino, B.~G. Janesko, R.~Gomperts,
  B.~Mennucci, H.~P. Hratchian, J.~V. Ortiz, A.~F. Izmaylov, J.~L. Sonnenberg,
  D.~Williams-Young, F.~Ding, F.~Lipparini, F.~Egidi, J.~Goings, B.~Peng,
  A.~Petrone, T.~Henderson, D.~Ranasinghe, V.~G. Zakrzewski, J.~Gao, N.~Rega,
  G.~Zheng, W.~Liang, M.~Hada, M.~Ehara, K.~Toyota, R.~Fukuda, J.~Hasegawa,
  M.~Ishida, T.~Nakajima, Y.~Honda, O.~Kitao, H.~Nakai, T.~Vreven,
  K.~Throssell, J.~A. Montgomery, {Jr.}, J.~E. Peralta, F.~Ogliaro, M.~J.
  Bearpark, J.~J. Heyd, E.~N. Brothers, K.~N. Kudin, V.~N. Staroverov, T.~A.
  Keith, R.~Kobayashi, J.~Normand, K.~Raghavachari, A.~P. Rendell, J.~C.
  Burant, S.~S. Iyengar, J.~Tomasi, M.~Cossi, J.~M. Millam, M.~Klene, C.~Adamo,
  R.~Cammi, J.~W. Ochterski, R.~L. Martin, K.~Morokuma, O.~Farkas, J.~B.
  Foresman, D.~J. Fox, Gaussian 16 {R}evision {C}.01, gaussian Inc. Wallingford
  CT (2016).

\bibitem{levy1991gaussian}
R.~M. Levy, M.~Belhadj, D.~B. Kitchen, Gaussian fluctuation formula for
  electrostatic free-energy changes in solution, J. Chem. Phys. 95~(5) (1991)
  3627--3633.

\bibitem{lorentz1881ueber}
H.~Lorentz, {U}eber die {A}nwendung des {S}atzes vom {V}irial in der
  kinetischen {T}heorie der {G}ase, Ann. phys. 248~(1) (1881) 127--136.

\bibitem{hamelin1982crystallographic}
A.~Hamelin, The crystallographic orientation of gold surfaces at the
  gold-aqueous solution interphases, J. Electroanal. Chem. Interf. Electrochem.
  142~(1-2) (1982) 299--316.

\bibitem{neugebauer1992adsorbate}
J.~Neugebauer, M.~Scheffler, Adsorbate-substrate and adsorbate-adsorbate
  interactions of {N}a and {K} adlayers on {A}l(111), Phys. Rev. B 46~(24)
  (1992) 16067.

\bibitem{polatoglou1993vacancy}
H.~Polatoglou, M.~Methfessel, M.~Scheffler, Vacancy-formation energies at the
  (111) surface and in bulk {A}l, {C}u, {A}g, and {R}h, Phys. Rev. B 48~(3)
  (1993) 1877.

\bibitem{stumpf1996ab}
R.~Stumpf, M.~Scheffler, Ab initio calculations of energies and self-diffusion
  on flat and stepped surfaces of {A}l and their implications on crystal
  growth, Phys. Rev. B 53~(8) (1996) 4958.

\bibitem{zhu2010corrosion}
J.~Zhu, L.~Hihara, Corrosion of continuous alumina-fibre reinforced {A}l--2
  wt.\% {C}u--{T}6 metal--matrix composite in 3.15 wt.\% {N}a{C}l solution,
  Corros. Sci. 52~(2) (2010) 406--415.

\end{thebibliography}

\end{document}